\documentclass[twocolumn,english,aps,prl,reprint,floatfix,notitlepage,footinbib,preprintnumbers,superscriptaddress]{revtex4-1}
\pdfoutput=1
\usepackage{lmodern}

\usepackage[T1]{fontenc}
\usepackage[latin9]{inputenc}
\usepackage{geometry}
\usepackage{color}
\usepackage{dsfont}
\usepackage{slashed}
\usepackage{ragged2e}
\usepackage{comment}
\usepackage{natbib}
\usepackage{babel}
\usepackage{amsmath,amssymb,amsthm,enumitem}
\usepackage{graphicx}
\usepackage{stackengine}
\usepackage{esint}
\usepackage[framemethod=TikZ]{mdframed}
\usepackage[unicode=true,pdfusetitle,
 bookmarks=true,bookmarksnumbered=false,bookmarksopen=false,
 breaklinks=false,pdfborder={0 0 1},backref=false,colorlinks=true]
 {hyperref}
\hypersetup{pdftitle={Signs, Spin, SMEFT: Sum Rules at Dimension Six},
 pdfauthor={Grant N. Remmen and Nicholas L. Rodd},
 citecolor=blue,linkcolor=black,urlcolor=black}
\usepackage{breakurl}
\usepackage[hang,flushmargin]{footmisc} 
\allowdisplaybreaks
\makeatletter

\usepackage{etoolbox}
\apptocmd{\thebibliography}{\justifying\setlength{\leftskip}{7.4mm}}{}{}

 \@ifundefined{textcolor}{}
 {%
   \definecolor{BLACK}{gray}{0}
   \definecolor{WHITE}{gray}{1}
   \definecolor{RED}{rgb}{1,0,0}
   \definecolor{GREEN}{rgb}{0,1,0}
   \definecolor{BLUE}{rgb}{0,0,1}
   \definecolor{CYAN}{cmyk}{1,0,0,0}
   \definecolor{MAGENTA}{cmyk}{0,1,0,0}
   \definecolor{YELLOW}{cmyk}{0,0,1,0}
 }

\usepackage{babel}

\makeatletter
\def\simgt{\mathrel{\lower2.5pt\vbox{\lineskip=0pt\baselineskip=0pt
           \hbox{$>$}\hbox{$\sim$}}}}
\def\simlt{\mathrel{\lower2.5pt\vbox{\lineskip=0pt\baselineskip=0pt
           \hbox{$<$}\hbox{$\sim$}}}}
\makeatother

\newcommand{\ket}[1]{\left| #1 \right\rangle}
\newcommand{\bra}[1]{\left\langle #1 \right |}

\newcommand{\be}{\begin{equation}}
\newcommand{\ee}{\end{equation}}

\newcommand{\Eq}[1]{Eq.~(\ref{#1})}
      
\begin{document}

\pagestyle{plain}

\title{Signs, Spin, SMEFT: Sum Rules at Dimension Six}

\author{Grant N. Remmen}
\affiliation{Kavli Institute for Theoretical Physics, University of California, Santa Barbara, CA 93106}
\thanks{e-mail: \url{remmen@kitp.ucsb.edu}, \url{nrodd@berkeley.edu}}
\affiliation{Department of Physics, University of California, Santa Barbara, CA 93106}
\author{Nicholas L. Rodd}
\affiliation{Berkeley Center for Theoretical Physics, University of California, Berkeley, CA 94720}
\thanks{e-mail: \url{remmen@kitp.ucsb.edu}, \url{nrodd@berkeley.edu}}
\affiliation{Theoretical Physics Group, Lawrence Berkeley National Laboratory, Berkeley, CA 94720}


\begin{abstract}
\noindent We place theoretical constraints on the leading deviations to four-fermion standard model interactions. 
Invoking S-matrix analyticity and partial wave unitarity, we develop new dispersion relations that yield either spin-dependent sum rules on dimension-six fermionic operators or information about the amplitude's behavior at large momentum. 
The pattern of SMEFT inequalities we find for theories obeying certain large-momentum constraints enables a diagnosis of properties of emerging new physics. 
These relations form a bridge between new physics searches: discovery of flavor-violating $\tau$ decays at Belle~II would motivate the search for flavor-conserving new physics below $25$~TeV, as the results would provide definitive information about the UV.
\end{abstract}
\maketitle

\noindent{\it Introduction.}---Testing the effective field theory (EFT) of the Standard Model (SM) is of increasing experimental and theoretical importance.
Prior to production of new on-shell states, the first observed signals of new physics will likely be subtle deviations from the SM emerging at the energy or precision frontier.
In the Wilsonian approach, these deviations are packaged in the SMEFT~\cite{Weinberg:1979sa,WarsawBasis,Alonso:2013hga,Lehman:2014jma,Lehman:2015coa,Henning:2015alf,Murphy:2020rsh,Li:2020gnx}, which systematically enumerates the operators of higher mass dimension built from SM fields, with coefficients dictated by the UV completion.
The quest for new physics is then reformulated as determining the signs and magnitudes of the SMEFT couplings or Wilson coefficients.

Given an EFT, one might expect that arbitrary values of the Wilson coefficients are allowed, but this is false.
Fundamental principles of quantum field theory, such as unitarity and locality of scattering amplitudes, constrain the space of coefficients~\cite{IRUV,Pham:1985cr,Ananthanarayan:1994hf,Pennington:1994kc}.
However, despite progress in constraining various EFTs~\footnote{For a broader review of the development of EFT bounds see, for example, Ref.~\cite{SMEFTbosons} and references therein.}, IR consistency techniques are only beginning to be systematically deployed for the full SMEFT, e.g., Refs.~\cite{SMEFTbosons,SMEFTfermions,Bellazzini:2018paj,Zhang:2018shp,Bi:2019phv,Wang:2020jxr,Zhang:2020jyn,Fuks:2020ujk,Yamashita:2020gtt,Azatov:2021ygj}.
A fundamental challenge is that power counting implies the presence of an obstacle in applying IR consistency to bound the SMEFT.
Beyond the dimension-five Weinberg operator, the most relevant SMEFT operators---and those of most immediate phenomenological importance---appear at mass dimension six, with basis enumerated in Ref.~\cite{WarsawBasis}.
Quartic dimension-six operators have amplitudes scaling as $p^2\sim s,t,u$.
But the optical theorem, which connects amplitudes to positive definite cross sections, applies to the forward limit: the dimension-six amplitude either vanishes as $t\rightarrow 0$ or flips sign under crossing symmetry $s\rightarrow -s$ (suppressing finite-mass corrections), obstructing a would-be bound without invoking additional assumptions~\cite{Donal}.
Extant positivity bounds thus typically apply to operators with at least four powers of momentum and hence SMEFT dimension eight or higher, while sign-agnostic sum rules do not always allow the extraction of detailed information about the characteristics (e.g., spin) of states in the UV.

In this paper, we connect the UV and IR to find new ways to constrain dimension-six fermionic operators in the form of sum rules.
We do so by making use of spin in a crucial way, performing a generalized spinning partial wave expansion to derive a condition that holds explicitly for fermionic amplitudes.
Combining our result with bedrock assumptions on the UV physics---namely, that it is Lorentz invariant, unitary, and local---we establish sum rule bounds for the dimension-six fermionic SMEFT that provide a powerful connection between the deep-UV momentum scaling of the amplitude, the spin of states in the completion of the SMEFT, and the low-energy Wilson coefficients.

Consistent with axioms of quantum field theory, we show that one of two possibilities must hold: either the physics generating dimension-six operators at low energies must produce amplitudes scaling as $p^2$ at generic scattering angle (i.e., no enhanced perturbative unitarity at general kinematics) or only scalar and vector partial waves can contribute, in such a way as to dictate the sign of the dimension-six operator depending on which spin dominates.
We emphasize that our prediction that enhanced (i.e., super-Froissart~\cite{Davighi:2021osh}) UV scaling of the amplitude leads to a spin-limited partial wave expansion is the result of the dispersion relation calculation, which itself relies only on Lorentz invariance, unitarity, and analyticity; the restriction to scalar and vector currents and the enhanced UV behavior are not independent assumptions.
The appearance of additional requirements on the UV in order to write down sum rules is a universal feature of dimension-six dispersion relations~\cite{Davighi:2021osh,Donal,Gu:2020thj,Azatov:2021ygj}.
Moreover, the discovery of Wilson coefficients outside of the region predicted by the sum rules would itself be of phenomenological interest, as it would give us direct information about the UV momentum scaling that cannot be directly observed in the IR and hence would inform model building.
Indeed, we will provide an explicit example where this could be realized experimentally.
Our results extend the set of previously known SMEFT sum rules, a comprehensive list of which was recently determined in Ref.~\cite{Gu:2020thj}.

We begin our discussion with a derivation of the fermionic sum rules.
We explore our result in the context of the operator
\be
{\cal O}_{e} = -b_{mnpq}^e(\bar e_m \gamma_\mu e_n)(\bar e_p \gamma^\mu e_q),
\label{eq:Oe}
\ee
explicitly indexing the $N_f$ flavors, and demonstrate that the coefficient matrix $b^e_{mnpq}$ cannot be chosen arbitrarily.
We then apply our constraints to the remaining four-fermion operators in the SMEFT.
Finally, we will demonstrate that our bounds are satisfied in several UV completions and discuss the mapping of our sum rules onto the experimental landscape, highlighting in particular their ramifications for searches of flavor and CP violation.

\bigskip

\noindent{\it Analyticity and Partial Wave Unitarity.}---We wish to use unitarity and the analytic properties of scattering amplitudes to constrain the Wilson coefficients of operators of the form given in \Eq{eq:Oe}.
By power counting, four-fermion scattering mediated by such operators will result in amplitudes $\propto p^2$.
Crossing-symmetric combinations will thus scale as $t$ and vanish in the forward limit.
Extracting the $\propto s$ part of a forward amplitude via a contour integral, one finds that, by analyticity,
\be
\begin{aligned}
&\lim_{s\rightarrow 0}\partial_s {\cal A}_{\alpha\beta}(s,0) = \frac{1}{2\pi i} \oint_\gamma \frac{{\rm d} s}{s^2} {\cal A}_{\alpha\beta}(s,0)\\
&= \frac{1}{\pi} \int_{s_0}^\infty \frac{{\rm d}s}{s^2} \left[{\rm Im}{\cal A}_{\alpha\beta}(s,0) {-} {\rm Im}{\cal A}_{\bar\alpha\beta}(s,0)\right] - C^{(s)}_{\infty,\alpha\beta}.
\end{aligned}\label{eq:forward}
\ee
Here $\gamma$ is a contour around the origin, parametrically small in $|s|$, so that the amplitude can be evaluated in the EFT.
The result on the second line follows after deforming the contour out to large $|s|$, where $s_0$ is the characteristic mass scale of the UV completion.
In the full amplitude, there are also massless branch cuts from SM fields extending all the way to $s=0$.
However, these can be ignored since we can always IR-regulate these cuts by turning on a small mass~\cite{IRUV,SMEFTbosons} or invoke a weak-coupling assumption to drop these massless loops as in Ref.~\cite{Nicolis:2009qm}.
The term $C^{(s)}_{\infty,\alpha\beta}$ is the contribution from the residue at infinity, 
\be
C_{\infty,\alpha\beta}^{(s)} {=}  {\rm Res}\left[\frac{{\cal A}_{\alpha\beta}(s,0)}{s^2},s{=}\infty\right]{=}  \, -\lim_{|s|\rightarrow\infty}\!\! \frac{{\cal A}_{\alpha\beta}(s,0)}{s}.
\ee
Note the Froissart bound~\cite{Froissart:1961ux,Martin:1962rt} constrains the asymptotic scaling of the amplitude by $\lim_{|s|\rightarrow\infty}|{\cal A}_{\alpha\beta}(s,0)| < O(s \log^2 s)$. 
The labels $\alpha,\beta$ represent scattered particles' quantum numbers, including their helicity, and $\bar\alpha$, $\bar\beta$ their conjugates.
(While helicity in general transforms nontrivially under crossing symmetry, in the forward limit it transforms just like an internal quantum number, e.g., flavor~\cite{Bellazzini:2016xrt}.)
We will consider elastic scattering amplitudes in this work, and therefore it is sufficient to state the quantum numbers of our initial state, which we take to be $\ket{\alpha,\beta}$.
Since we expect the new physics to be at least above the weak scale, throughout we take the fermions to be effectively massless.
The optical theorem then implies ${\rm Im}\,{\cal A}(s,0) = s\,\sigma(s)>0$, so the integrand in \Eq{eq:forward} is of indefinite sign, precluding a positivity bound from the forward amplitude alone; one can nonetheless use \Eq{eq:forward} to derive sum rules on dimension-six operators, relating the Wilson coefficients to a difference of UV cross sections~\cite{Wang:2020jxr,Low:2009di,Falkowski:2012vh,Bellazzini:2014waa,Ema:2018jgc,Gu:2020thj,Azatov:2021ygj}.

Before proceeding, we briefly note a technical detail generally associated with dispersion relations similar to \Eq{eq:forward}. 
The goal of such dispersion relations is to isolate the physics associated with quartic dimension-six operators, which scale as $s$ or $t$.
Yet in the SM, the exchange of massless particles, such as the photon, can lead to amplitudes scaling as $s/t$.
Introducing an IR mass regulator, these singularities will produce finite and matching contributions on both sides of our dispersion relations, on the left arising from the contour around the origin, while on the right originating from the contour at infinity.
Accordingly, we can formally remove the poles from both sides to isolate the contributions of interest, thereby considering a pole-subtracted dispersion relation~\footnote{In contrast, such a procedure would not be valid to remove the $t$-channel singularity in gravity, unless additional assumptions were invoked, because, unlike quantum electrodynamics, the EFT of general relativity is nonrenormalizable on its own, and thus will exhibit unitarizing UV states in the complex plane} (for additional discussion of this point, see, e.g., Refs.~\cite{deRham:2017imi,Gu:2020thj}).
The necessity of this subtraction is a general feature of dispersive dimension-six arguments, e.g., conventional sum rules~\cite{Low:2009di,Falkowski:2012vh,Bellazzini:2014waa,Gu:2020thj,Azatov:2021ygj}, and not unique to the results derived in this paper.
Moreover, since we will be working at leading order in the dimension-six SMEFT Wilson coefficients, no higher-dimension operators containing intermediate SMEFT bosons contribute to our four-fermion amplitudes~\footnote{The Weinberg operator or cubic boson operators like $W_{\mu\nu}^3$ can only contribute at loop level, while in our choice of basis~\cite{WarsawBasis} there are neither dimension-six SMEFT operators containing two bosons nor one boson and two fermions.}.

In order to extract more information from dispersion relations and thereby constrain the UV by what we observe in the IR, we can make use of more properties of an amplitude than its forward limit alone.
For scalar amplitudes, a partial wave expansion reveals that all $t$-derivatives of ${\rm Im}\,{\cal A}$ are positive in the forward limit, as a consequence of properties of derivatives of Legendre polynomials~\cite{Nicolis:2009qm}.
It is therefore well motivated to extract the $t$-derivative of the amplitude; performing an unsubtracted dispersion relation and again taking the forward limit so that helicity transforms simply under crossing, we have~\footnote{We can apply this dispersion relation when $\cos\theta$ lies within the Martin-Lehmann ellipse; see Refs.~\cite{Lehmann:1958ita,Martin:1965jj,Nicolis:2009qm}.
When massless loops are included, the ellipse is just the real interval $[-1,1]$, corresponding to physical $t=s(\cos\theta-1)/2$.
This requirement poses no problem for our argument, since the contour at the origin extracting ${\cal A}(0,t)$ can instead be taken at small, finite $s$ well below the UV cutoff, allowing $t$ to be finite and negative; in any case, all of our dispersion relations ultimately take the forward limit $t\rightarrow 0$.}:
\be 
\begin{aligned}
&\lim_{t\rightarrow 0} \partial_t {\cal A}_{\alpha\beta}(0,\! t) = - C^{(t)}_{\infty,\alpha\beta} \\
&+\!\frac{1}{\pi}\! \int_{s_0}^\infty \!\!\!\!{\rm d}s \lim_{t\rightarrow 0} \partial_t \!\left[\!\frac{{\rm Im}{\cal A}'_{\alpha\beta}(s,\! t)}{s}\! +\! \frac{{\rm Im}{\cal A}'_{\bar\alpha \beta}(s,\! t)}{s+t}\!\right],\label{eq:generalt}
\end{aligned}
\ee
where the boundary in this case is given by another residue at infinity,
\be 
C_{\infty,\alpha\beta}^{(t)} = {\rm Res}\left[\frac{\lim_{t\rightarrow 0} \partial_t {\cal A}'_{\alpha\beta}(s,t)}{s},s=\infty\right].
\ee

In \Eq{eq:generalt}, the prime appearing on all amplitudes on the right-hand side is used to signify the presence of an additional sign that must be accounted for.
The sign results from fermion interchange, and explicitly arises as we have taken $s \to 0$ for $t \neq 0$ on the left-hand side.
In order to explain this explicitly, let us outline the conventions we will adopt throughout this work.
We use mostly-plus metric signature, take external momenta to be labeled cyclicly, and treat the initial and final states as incoming and outgoing, respectively.
The independent Mandelstam invariants are then given by $s=-(p_1+p_2)^2$ and $t=-(p_1-p_4)^2$.
As we consider elastic scattering, our initial and final states are given by $\ket{i} = \ket{\alpha,\beta}$ and $\bra{f} = \bra{\bar\beta,\bar\alpha}$.
Accordingly, we can see that the kinematic limit taken in \Eq{eq:forward} is straightforward: $t \to 0$ with $s \neq 0$ amounts to the limit of forward scattering $p_4 \to p_1$ and $p_3 \to p_2$.
In contrast, taking $s \to 0$ with $t \neq 0$, as we do on the left of \Eq{eq:generalt}, forces $p_2 \to - p_1$ and $p_3 \to - p_4$.
The signs physically imply that particle 2, previously incoming, is now outgoing, and vice versa for particle 3.
Particles 2 and 3 have therefore been crossed, which consequently introduces an overall sign as they are fermions.
We can see this explicitly by considering an example of the form that the four-fermion amplitude can take.
Written in terms of spinor-helicity variables, if we scatter states with opposite helicity, the little-group scaling implies that the amplitude must be proportional to $[13]\langle 24\rangle$.
Taking the limit of forward scattering ($t \to 0$ with $s \neq 0$), we have $[13]\langle 24\rangle \to [12] \langle 21 \rangle = s$.
However, if we take $s \to 0$ with $t \neq 0$, then $[13]\langle 24\rangle \to [1 \,{-}4] \langle {-}1\,4 \rangle = - [14] \langle 14 \rangle = -t$.
The additional sign that arises here originates from the canonical analytic continuation for spinors, $|\!-\!p\rangle = -|p\rangle$ and $|\!-\!p]=|p]$ (see, e.g., Ref.~\cite{EH}).
The prime in \Eq{eq:generalt} is introduced to track that this crossing has occurred, and explicitly we have ${\cal A}'_{\alpha\beta}=-{\cal A}_{\alpha\beta}$ and ${\cal A}'_{\bar\alpha\beta}=-{\cal A}_{\bar\alpha\beta}$.

Let us now perform a partial wave expansion of the two dispersion relations.
In the case where we are scattering particles of opposite helicity, so that $\alpha$ ($\beta$) corresponds to helicity $\pm 1/2$ ($\mp 1/2$), the initial state has total spin $1$.
For such a state, the Jacob-Wick expansion allows us to write the amplitude in terms of partial waves of definite total angular momentum, via a spinning generalization of the Legendre polynomials~\cite{Jacob:1959at,Horejsi:1993hz,ItzyksonZuber,Arkani-Hamed:2017jhn},
\be 
{\cal A}_{\alpha\beta}(s,t)=16\pi \sum_{j=1}^\infty (2j+1)a^{(j)}_{\alpha\beta}(s) d^{(j)}_{11}(\cos\theta),
\ee 
where the sum in $j$ starts at $1$ because an $s$-wave coupling to our spin-one initial state is forbidden by angular momentum conservation; similar selection rules for four-fermion SMEFT operators and their implications for collider physics are discussed in Ref.~\cite{Alioli:2020kez}.
Here, $d^j_{mm'}$ is the Wigner small $d$-matrix~\cite{Wigner,Horejsi:1993hz}, which for $m\,{=}\,m'\,{=}\,1$ can be written compactly as $d_{11}^{(j)}(x)= (1+x)P_{j-1}^{(0,2)}(x)/2$, where $P_j^{(m,n)}$ are the Jacobi polynomials~\footnote{More generally, the partial wave expansion in the total angular momentum basis is performed with the Wigner $D$-matrices, irreducible representations of ${\rm SU}(2)$, but these only differ from the small $d$-matrices by a little group phase factor, which can be set to unity by an appropriate choice of coordinates}.
Meanwhile, the amplitude ${\cal A}_{\bar\alpha\beta}$ has zero net helicity and can thus be expanded in terms of the Legendre polynomials $P_j = P_j^{(0,0)}$,
\be
{\cal A}_{\bar{\alpha}\beta}(s,t)=16\pi \sum_{j=0}^\infty (2j+1)a^{(j)}_{\bar\alpha\beta}(s) P_j(\cos\theta),
\ee
where the sum here starts at zero angular momentum.
Exchanging differentiation in $t$ for $\cos\theta=1+2t/s$ and keeping track of signs from little group phases, Eqs.~\eqref{eq:forward} and \eqref{eq:generalt} become
\begin{widetext}
\be
\begin{aligned}
&\lim_{s\rightarrow0}\partial_{s}{\cal A}_{\alpha\beta}(s,0) + C_{\infty,\alpha\beta}^{(s)}&&\hspace{-1em} = -\lim_{s\rightarrow0}\partial_{s}{\cal A}_{\bar\alpha\beta}(s,0) - C_{\infty,\bar\alpha\beta}^{(s)}
\\&&&\hspace{-1em} =16\int_{s_{0}}^{\infty}\frac{{\rm d}s}{s^{2}}\left\{ -{\rm Im}a_{\bar\alpha\beta}^{(0)}(s)+\sum_{j=1}^{\infty}(2j+1)\left[{\rm Im}a_{\alpha\beta}^{(j)}(s)-{\rm Im}a_{\bar\alpha\beta}^{(j)}(s)\right]\right\}\\
&{\rm and}\vspace{5mm} \\
&\lim_{t\rightarrow0}\partial_{t}{\cal A}_{\alpha\beta}(0,t) +C_{\infty,\alpha\beta}^{(t)}&&\hspace{-1em}= \lim_{t\rightarrow0}\partial_{t}{\cal A}_{\bar\alpha\beta}(0,t)+\lim_{s\rightarrow0}\partial_{s}{\cal A}_{\bar\alpha\beta}(s,0) +C_{\infty,\bar\alpha\beta}^{(s)} + C_{\infty,\bar\alpha\beta}^{(t)}
\\&&&\hspace{-1em} =16\int_{s_{0}}^{\infty}\frac{{\rm d}s}{s^{2}}\left\{+{\rm Im}a_{\bar\alpha\beta}^{(0)}(s)-\sum_{j=1}^{\infty}(2j+1)\left[j(j+1)-1\right]\left[{\rm Im}a_{\bar\alpha\beta}^{(j)}(s)+{\rm Im}a_{\alpha\beta}^{(j)}(s)\right]\right\}.\label{eq:dispbig}
\end{aligned}
\ee
\newpage
\end{widetext}

\noindent Adding these two dispersion relations, we have 
\be 
\begin{aligned}
&\lim_{t\rightarrow0}\partial_{t}{\cal A}_{\alpha\beta}(0,t){+}\lim_{s\rightarrow0}\partial_{s}{\cal A}_{\alpha\beta}(s,0){+}C_{\infty,\alpha\beta}^{(s)}{+}C_{\infty,\alpha\beta}^{(t)} 
\\&=\lim_{t\rightarrow0}\partial_{t}{\cal A}_{\bar\alpha\beta}(0,t) + C_{\infty,\bar\alpha\beta}^{(t)} 
\\& =-16\int_{s_{0}}^{\infty}\frac{{\rm d}s}{s^{2}}\sum_{j=1}^{\infty}\left\{(2j+1)j(j+1){\rm Im}a_{\bar\alpha\beta}^{(j)}(s)\right. \\& \qquad\qquad\qquad \left.+(2j+1)\left[j(j+1)-2\right]{\rm Im}a_{\alpha\beta}^{(j)}(s)\right\}.
\label{eq:disp}
\end{aligned}
\ee

By partial wave unitarity, the imaginary parts of the partial waves appearing on the right-hand sides of the dispersion relations in Eqs.~\eqref{eq:dispbig} and \eqref{eq:disp} are all nonnegative.
Since the right-hand side of \Eq{eq:disp} is less than or equal to zero, if $C_{\infty,\alpha\beta}^{(s)} + C_{\infty,\alpha\beta}^{(t)}$ or $C_{\infty,\bar\alpha\beta}^{(t)}$ vanish, one could a priori anticipate a bounded sign on the Wilson coefficients.
We do not assert that these boundary conditions vanish in general however, and if the sign predicted in this case were experimentally found to be violated, the dispersion relation in turn could be used to gain information about the UV scaling of the amplitude (i.e., the size/sign of the boundary terms) by virtue of IR measurements.
Moreover, since the EFT amplitude for opposite-helicity scattering is constrained to be proportional to $[13]\langle24\rangle$, we will always have $\lim_{t\rightarrow0}\partial_{t}{\cal A}_{\alpha\beta}(0,t)+\lim_{s\rightarrow0}\partial_{s}{\cal A}_{\alpha\beta}(s,0)=0$ (and similarly, ${\cal A}_{\bar\alpha\beta}(0,t)=0$). 
By the dispersion relation in Eq.~\eqref{eq:disp}, if the boundary terms vanish, then we must have vanishing imaginary parts of the partial waves $a_{\alpha\beta}^{(j)}$ and $a_{\bar\alpha\beta}^{(j)}$ for all $j\geq2$ (and ${\rm Im}\,a^{(1)}_{\bar\alpha\beta}(s)=0$ as well).
That is, only scalar and vector channels are allowed to contribute to the cross section.
Note that we have not assumed this restriction on angular momentum as input; rather, it is an output of bedrock QFT axioms, plus a condition on the UV amplitude.
If higher-$j$ channels are nonzero, for either amplitude, analyticity implies that the boundary term(s) cannot vanish (or that the partial wave expansion does not converge, which occurs if ${\rm Im}\,a \gtrsim 1/j^3$ asymptotically).
The boundary terms vanish if the amplitude obeys super-Froissart conditions~\cite{Davighi:2021osh}, $\lim_{|s|\rightarrow\infty}|{\cal A}_{\alpha\beta}(s,0)| < O(s)$ and $\lim_{|s|\rightarrow\infty} \lim_{t\rightarrow 0} \partial_t|{\cal A}_{\alpha\beta}(s,t)|< O(s^0)$.
While boundary terms do not vanish for all theories, one can view this as a predictive feature of our dispersion relations: in either case, our results effectively function as phenomenological diagnostics of the characteristics of new physics based on future observation of the sign and/or relative magnitude of higher-dimension operators.
The residue at infinity and the details of the derivative partial wave expansion will be explored further in Ref.~\cite{upcoming} in a broader context for the full dimension-six SMEFT that falls outside the scope of the present work.

If we are in the case in which the boundary terms vanish (and the partial wave expansion converges), then for opposite-helicity fermion scattering, analyticity and unitarity imply that the partial wave expansion contains only $j=0,1$ terms, and we have: 
\begin{mdframed}[linewidth=1.5pt, roundcorner=10pt]
\vspace{-10pt}
\be 
\begin{aligned}
& \lim_{s\rightarrow0}\partial_{s}{\cal A}_{\bar\alpha\beta}(s,0) = -\lim_{s\rightarrow0}\partial_{s}{\cal A}_{\alpha\beta}(s,0) \hspace{-8pt} \\&=16\int_{s_{0}}^{\infty}\frac{{\rm d}s}{s^{2}}\left[{\rm Im}a_{\bar\alpha\beta}^{(0)}(s)-3\,{\rm Im}a^{(1)}_{\alpha\beta}(s)\right]. \label{eq:disprestricted}
\end{aligned}
\ee
\end{mdframed}
As both ${\rm Im}a_{\bar\alpha\beta}^{(0)}(s)$ and ${\rm Im}a^{(1)}_{\alpha\beta}(s)$ are positive, the sign of this expression indicates which states are dominating in the UV; if the completion is dominated by scalars, then the integral is positive, whereas if it is dominated by vectors, the result will be negative.
We will next determine the implications of this result for the SMEFT.

\bigskip

\noindent{\it Dimension-Six Bounds.}---Let us first apply the sum rule in \Eq{eq:disprestricted} to the operator in \Eq{eq:Oe}.
In detail, through the operator we scatter a superposition of right-handed electron flavors weighted by $\alpha_m$, $\beta_m$, $\gamma_m$, $\delta_m$ for particles 1, 2, 3, 4, respectively.
In spinor-helicity notation, we find~\footnote{Care must be taken not just for the relative sign between channels, but for the overall sign of the amplitude.
The relative order of the initial and final two-particle states can be chosen freely, but it is convenient to define the final state such that in the elastic forward limit we have ${\bra{f}}= {\ket{i}}^{\dagger}$, as this allows for a straightforward application of the optical theorem.
To be explicit, the conventions we establish in the text imply that when scattering opposite-helicity states, we take ${\ket{i}} = {\ket{1,2}}$ and ${\bra{f}} = {\bra{3,4}} = -{\bra{4,3}}$.
With the external states as defined, before performing any Wick contractions, the amplitude takes the following form (suppressing flavor):
\begin{equation*}
\begin{aligned}
{\langle} 0 {\vert}
a_{p_3} b_{p_4}
(\bar{e} \gamma^{\mu} e) 
(\bar{e} \gamma_{\mu} e) 
b^{\dagger}_{p_1} a^{\dagger}_{p_2}
{\vert} 0 {\rangle},
\end{aligned}
\end{equation*}
where $a^{\dagger}$ and $b^{\dagger}$ are the particles' and antiparticles' creation operators, respectively.
There are four possible contractions.
The contraction $(\bar{e} \gamma^{\mu} e) (\bar{e} \gamma_{\mu} e) \to (\bar{1} 2)(\bar{3} 4)$ generates no relative minus sign, nor does $(\bar{3} 4) (\bar{1} 2)$.
The two contractions $(\bar{3} 2)(\bar{1} 4)$ and $(\bar{1} 4)(\bar{3} 2)$, however, will.
Consequently, the full amplitude is given by
\begin{equation*}
\begin{aligned}
&\hspace{5mm}&&{\cal A}(\bar{e}^{-}e^{+}\to e^{+}\bar{e}^{-})
\\&&&= -2\alpha_{m}\beta_{n}\gamma_{p}^*\delta_{q}^*b^e_{mnpq}({[1 \vert \gamma_{\mu} \vert 2\rangle} {[3 \vert \gamma^{\mu} \vert 4\rangle} 
\\&&&\hspace{32mm} - {[3 \vert \gamma_{\mu} \vert 2\rangle} {[1 \vert \gamma^{\mu} \vert 4\rangle})
\\&&&=-8\alpha_{m}\beta_{n}\gamma_{p}^*\delta_{q}^*b^e_{mnpq} {[13]} {\langle 24\rangle},
\end{aligned}
\end{equation*}
as given in \Eq{eq:e4amp}}:
\be 
{\cal A}(\bar{e}^{-}e^{+}\to e^{+}\bar{e}^{-})=-8\alpha_{m}\beta_{n}\gamma_{p}^*\delta_{q}^*b^e_{mnpq} [13] \langle24\rangle .\label{eq:e4amp}
\ee
The helicities have been chosen in a forward configuration; similarly, we must take forward flavor vectors, $\delta_m = \alpha_m$ and $\gamma_m = \beta_m$.
When $t=0$, we have $p_4 = p_1$ and $p_3 = p_2$, so the amplitude becomes
\be 
{\cal A}(\bar{e}^{-}e^{+}\to e^{+}\bar{e}^{-})|_{t=0} =-8\alpha_{m}\beta_{n}\beta^*_{p}\alpha^*_{q}b^e_{mnpq}s.
\ee
Thus, for vanishing boundary terms, our sum rule in \Eq{eq:disprestricted} implies that the combination
\be 
b^e_{mnpq} \alpha_{m}\beta_{n}\beta^*_{p}\alpha^*_{q} \! = \! 2\!\!\int_{s_{0}}^{\infty}\!\!\frac{{\rm d}s}{s^{2}}\!\!\left[{\rm Im}a_{\bar\alpha\beta}^{(0)}(s){-}3{\rm Im}a^{(1)}_{\alpha\beta}(s)\right] \label{eq:e4bound}
\ee
is positive (respectively, negative) if the UV is dominated by scalars (vectors)~\footnote{A similar structure appears in the computation of the nucleon-nucleon scattering length in the ${}^1\! S_0$ singlet state with the operators $C_S(N^\dagger N)^2$ and $C_T (N^\dagger \boldsymbol{\sigma} N)^2$ replacing the scalar and vector channels, respectively~\cite{Mehen:1999qs,Donal}.}.

Similar bounds can be established for the remaining four-fermi SMEFT operators.
We consider all operators containing an even number of each type of SM fermion, allowing for two-to-two scattering for states of fixed SM charges.
There are two classes: the self-quartics,
\be 
\begin{aligned}
{\cal O}_{u} &= -b_{mnpq}^u(\bar u_m \gamma_\mu u_n)(\bar u_p \gamma^\mu u_q) \\
{\cal O}_{d} &= -b_{mnpq}^d(\bar d_m \gamma_\mu d_n)(\bar d_p \gamma^\mu d_q) \\
{\cal O}_{L} &= -b_{mnpq}^L(\bar L_m \gamma_\mu L_n)(\bar L_p \gamma^\mu L_q) \\
{\cal O}_{Q1} &=-b_{mnpq}^{Q1}(\bar Q_m \gamma_\mu Q_n)(\bar Q_p \gamma^\mu Q_q) \\
{\cal O}_{Q2} &= -b_{mnpq}^{Q2}(\bar Q_m \gamma_\mu \tau^I Q_n)(\bar Q_p \gamma^\mu \tau^I Q_q),
\end{aligned}\label{eq:O1}
\ee
together with \Eq{eq:Oe}, and the cross-quartics,
\be 
\begin{aligned}
{\cal O}_{eu} &= -c_{mnpq}^{eu}(\bar e_m \gamma_\mu e_n)(\bar u_p \gamma^\mu u_q) \\
{\cal O}_{ed} &= -c_{mnpq}^{ed}(\bar e_m \gamma_\mu e_n)(\bar d_p \gamma^\mu d_q) \\
{\cal O}_{ud1} &=-c_{mnpq}^{ud1}(\bar u_m \gamma_\mu u_n)(\bar d_p \gamma^\mu d_q) \\
{\cal O}_{ud2} &= -c_{mnpq}^{ud2}(\bar u_m \gamma_\mu T^a u_n)(\bar d_p \gamma^\mu T^a d_q) \\
{\cal O}_{LQ1} &= -c_{mnpq}^{LQ1}(\bar L_m \gamma_\mu L_n)(\bar Q_p \gamma^\mu Q_q) \\
{\cal O}_{LQ2} &= -c_{mnpq}^{LQ2}(\bar L_m \gamma_\mu \tau^I L_n)(\bar Q_p \gamma^\mu \tau^I Q_q)\\
{\cal O}_{Le} &= +c_{mnpq}^{Le}(\bar L_m \gamma_\mu L_n)(\bar e_p \gamma^\mu e_q) \\
{\cal O}_{Lu} &= +c_{mnpq}^{Lu}(\bar L_m \gamma_\mu L_n)(\bar u_p \gamma^\mu u_q) \\
{\cal O}_{Ld} &= +c_{mnpq}^{Ld}(\bar L_m \gamma_\mu L_n)(\bar d_p \gamma^\mu d_q) \\
{\cal O}_{Qe} &= +c_{mnpq}^{Qe}(\bar Q_m \gamma_\mu Q_n)(\bar e_p \gamma^\mu e_q) \\
{\cal O}_{Qu1} &= +c_{mnpq}^{Qu1}(\bar Q_m \gamma_\mu Q_n)(\bar u_p \gamma^\mu u_q) \\
{\cal O}_{Qu2} &= +c_{mnpq}^{Qu2}(\bar Q_m \gamma_\mu T^a Q_n)(\bar u_p \gamma^\mu T^a u_q) \\
{\cal O}_{Qd1} &= +c_{mnpq}^{Qd1}(\bar Q_m \gamma_\mu Q_n)(\bar d_p \gamma^\mu d_q) \\
{\cal O}_{Qd2} &= +c_{mnpq}^{Qd2}(\bar Q_m \gamma_\mu T^a Q_n)(\bar d_p \gamma^\mu T^a d_q).
\end{aligned}\label{eq:O2}
\ee
The ${\rm SU}(N)$ generators are $\tau^I = \sigma^I/2$ and $T^a = \lambda^a/2$, writing $\sigma^I$ and $\lambda^a$ for the Pauli and Gell-Mann matrices, respectively.
For consistency, we have defined the operators in \Eq{eq:O2} with a relative minus sign when the cross-quartic contains fermion species of opposite handedness, since we can introduce charge-conjugated fields to rewrite all such operators in terms of a single chirality, for example, $(\bar L_m \gamma_\mu L_n)(\bar e_p \gamma^\mu e_q) = - (\bar L_n^c \gamma_\mu L_m^c)(\bar e_p \gamma^\mu e_q)$ \footnote{Note that the sign difference between operators of identical and mixed chirality does not persist to the equivalent dimension-eight terms, which were considered in Ref.~\cite{SMEFTfermions}.
As an explicit example, in the notation of Ref.~\cite{SMEFTfermions}, up to a total derivative and terms that vanish on-shell, we have 
\begin{equation*}
\begin{aligned}
{\cal O}_{K1}[e,L] 
&= - a_{mnpq}^{eL,1} 
(\bar{e}_m \gamma_{\mu} \partial_{\nu} e_n)
(\bar{L}_p \gamma^{\nu} \partial^{\mu} L_q) \\
&= + a_{mnpq}^{eL,1} 
(\bar{e}_m \gamma_{\mu} \partial_{\nu} e_n)
(\partial^{\mu} \bar{L}_q^c \gamma^{\nu} L^c_p) \\
&= - a_{mnpq}^{eL,1} 
(\bar{e}_m \gamma_{\mu} \partial_{\nu} e_n)
(\bar{L}_q^c \gamma^{\nu} \partial^{\mu} L^c_p),
\end{aligned}
\end{equation*}
so that when the operator is written in terms of a single chirality, the same overall sign applies
}. 
Each of the $b$ tensors satisfies $b_{mnpq} = b_{pqmn}$ by symmetry and $b_{mnpq} = b_{qpnm}^*$ by hermiticity, and for $e^4$ we additionally have $b^e_{mnpq} = b^e_{pnmq}$ by a Fierz identity.
The $c$ tensors satisfy only the hermiticity property.
In total, the operators in Eqs.~\eqref{eq:Oe}, \eqref{eq:O1}, and \eqref{eq:O2} have $N_f^2(67N_f^2+2N_f+11)/4$ independent real coefficients, i.e., 1395 in the case $N_f = 3$ (786 CP-even and 609 CP-odd parameters)~\cite{Alonso:2013hga}.
This list does not exhaust the four-fermi operators at dimension six, as there are also baryon- and lepton-number violating operators in the SMEFT~\cite{WarsawBasis}.
In this work, however, we will restrict consideration to scattering of fixed SM charges, so these operators will not contribute; we leave consideration of superpositions of charge eigenstates---where these additional operators would be required and potentially bounded---to future work.

To exemplify how the argument proceeds for additional operators, for $L^4$, where we have both flavors and ${\rm SU}(2)$ charges, the amplitude ${\cal A}(\bar{L}^{+}L^{-} \to L^{-}\bar{L}^{+})$ is given by $-4b^{L}_{mnpq}(\alpha_{mi}\beta_{ni}\gamma_{pj}\delta_{qj}+\alpha_{mi}\beta_{qj}\gamma_{pj}\delta_{ni})\langle 13\rangle[24]$.
Marginalizing over the ${\rm SU}(2)$ charges results in two independent sum rules of the form in \Eq{eq:e4bound} but with $b_{mnpq}^e \alpha_m \beta_n \beta_p^* \alpha_q^*$ replaced by the two combinations:
\be
\begin{aligned}
&\tfrac{1}{2}b^{L}_{mnpq}\left(\alpha_{m}\beta_{n}\beta_{p}^{*}\alpha_{q}^{*}+\alpha_{m}\alpha_{n}^{*}\beta_{p}^{*}\beta_{q}\right) \\
&\tfrac{1}{2}b^{L}_{mnpq}\alpha_{m}\alpha_{n}^{*}\beta_{p}^{*}\beta_{q}.
\end{aligned}\label{eq:L4bounds}
\ee 
For all of the self-quartics, we define $b_{mnpq}^{\pm}=(b_{mnpq}\pm b_{mqpn})/2$ and $b_{mnpq}^{\pm}\alpha_{m}\alpha_{n}^{*}\beta_{p}^{*}\beta_{q}=b_{\alpha\beta}^{\pm}$.
The full set of operators appearing on the left-hand sides of our sum rules can thus be expressed compactly as
\be
\begin{aligned}
b_{\alpha\beta}^{e+},   &\;\;\;\qquad && \tfrac{1}{2}b_{\alpha\beta}^{L+}+ \tfrac{1}{2}b_{\alpha\beta}^{L-} ,&  \\
b_{\alpha\beta}^{L+},   &&& \tfrac{1}{2}b_{\alpha\beta}^{u+}+\tfrac{1}{2}b_{\alpha\beta}^{u-}, &  \\
b_{\alpha\beta}^{u+},   &&& \tfrac{1}{2}b_{\alpha\beta}^{d+}+\tfrac{1}{2}b_{\alpha\beta}^{d-}, &\\
b_{\alpha\beta}^{d+},   &&& b_{\alpha\beta}^{Q1+} + \tfrac{1}{4}b_{\alpha\beta}^{Q2+}, & \\
& \hspace{-8.5mm}\rlap{$\tfrac{1}{2}b_{\alpha\beta}^{Q1+}+\tfrac{1}{2}b_{\alpha\beta}^{Q1-} + \tfrac{1}{8}b_{\alpha\beta}^{Q2+} -\tfrac{3}{8}b_{\alpha\beta}^{Q2-} $,}
\end{aligned}\label{eq:abounds}
\ee
where the right-hand side of the sum rule is as in \Eq{eq:e4bound} for the analogous scattering process.

We can bound the cross-quartic operators in \Eq{eq:O2} in the same way.
For example, ${\cal A}(\bar{e}^{-}u^{+} \to u^{+}\bar{e}^{-}) = -2c_{mnpq}^{eu}\alpha_{m}\beta_{qi}\gamma_{pi}^*\delta_{n}^*[13]\langle 24\rangle$, implying the sum rule of the form in \Eq{eq:e4bound} with $c^{eu}_{mnpq}\alpha_{m}\alpha_{n}^{*}\beta_{p}^{*}\beta_{q}/4$ appearing on the left-hand side.
For each of the $c$ tensors, we now define $c_{mnpq} \alpha_{m}\alpha_{n}^{*}\beta_{p}^{*}\beta_{q} = c_{\alpha\beta}$, in terms of which the full set of operators appearing in the sum rules can be summarized as
\be 
\begin{aligned}&\tfrac{1}{4}c_{\alpha\beta}^{eu} &  &\;\;\;&& \tfrac{1}{4}c_{\alpha\beta}^{Le} & \\
&\tfrac{1}{4}c_{\alpha\beta}^{ed} &  &&&  \tfrac{1}{4}c_{\alpha\beta}^{Lu}& \\
&\tfrac{1}{4} c_{\alpha\beta}^{ud1}+ \tfrac{1\pm 3}{48} c_{\alpha\beta}^{ud2}  &  &&& \tfrac{1}{4}c_{\alpha\beta}^{Ld} &   \\
&\tfrac{1}{4}c_{\alpha\beta}^{LQ1}\pm \tfrac{1}{16} c_{\alpha\beta}^{LQ2}  &  &&& \tfrac{1}{4} c^{Qe}_{\alpha\beta} & \\
&&&&&   \tfrac{1}{4} c_{\alpha\beta}^{Qu1}+ \tfrac{1\pm 3}{48} c_{\alpha\beta}^{Qu2} &   \\ &&&&& \tfrac{1}{4} c_{\alpha\beta}^{Qd1}+ \tfrac{1\pm 3}{48} c_{\alpha\beta}^{Qd2}.&
\end{aligned}\label{eq:cbounds}
\ee
Recall that we use mostly-plus metric in our sum rules' relations between the operators in Eqs.~\eqref{eq:abounds} and \eqref{eq:cbounds} and the spin of states in the completion.
For the opposite signature convention, flip the sign of the right-hand sides of all dispersion relations.
As we will see, if either sign dominates in the sum rule (i.e., if the overall completion is dominated by scalars or vectors), then similarly with Ref.~\cite{SMEFTfermions} we obtain a sum rule in which flavor- and CP-violating terms are upper-bounded in magnitude by their symmetry-conserving analogues.

\bigskip

\noindent{\it Discussion.}---We can test our sum rules and examine the detailed operation of our dispersion relations by considering several realistic example UV completions. 
Due to the complicated dependence on kinematics typical of loop diagrams, we generically expect loop-level completions to generate an infinite tower in the partial wave expansion and so, by virtue of our dispersion relations, necessarily generate nonzero boundary term(s); we leave the investigation of this possibility to future work.
For illustrative purposes, let us therefore construct example tree-level completions.
(Note that, while any interacting theory will possess loop diagrams, and hence generically a full tower of nonzero partial waves, for a weakly coupled theory whose leading diagrams are at tree level we can consistently truncate to the amplitude in both the UV and the IR and consistently apply the dispersion relation at that order in the coupling~\cite{IRUV,Nicolis:2009qm}.)
For a thorough discussion of tree-level completions of SMEFT operators at dimension six, see Ref.~\cite{deBlas:2017xtg}.
Consider a complex scalar singlet $\phi$ of mass $m_\phi$ coupled via $y_{mn}\phi (\bar e_m e_n^{\rm c}) + {\rm h.c}$.
Electron-positron scattering thus proceeds through the $u$-channel:
\begin{center}\vspace{-0.1cm}
\includegraphics[height=2cm]{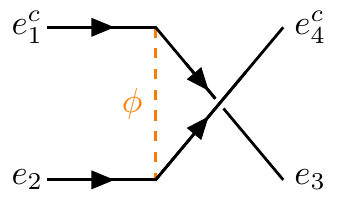}
\vspace{-0.1cm}\end{center}
and we have ${\cal A}_{\alpha\beta} = 4\alpha_m \beta_n \beta^*_p \alpha^*_q y_{mp}y^*_{nq}[13]\langle 24\rangle/(u-m_\phi^2 + i\epsilon)$.
This theory generates low-energy Wilson coefficients $b^{e+}_{\alpha\beta} = |y_{mn} \alpha_m\beta_n^*|^2/2m_\phi^2>0$.
For this completion, explicit computation shows that the boundary terms vanish, and positive $b^{e+}_{\alpha\beta}$ arising from a scalar current in the UV is consistent with our sum rule in \Eq{eq:e4bound}.
Since for this $u$-channel process ${\rm Im}\,a_{\alpha\beta}^{(j)} \propto \delta(s+m_\phi^2)$, the only contribution to the right-hand side of \Eq{eq:disp} comes from the crossed amplitude, ${\rm Im}\,a_{\bar\alpha\beta}^{(j)}(s) = \frac{1}{4} |y_{mn}\alpha_m \beta_n^*|^2\, s \,\delta_{j0} \,\delta(s{-}m_\phi^2)$.
Inputting this into the integrand, we find that all of our dispersion relations are satisfied.

Alternatively, by taking a complex vector $A^\mu$, a hypercharged, ${\rm SU}(2)$ doublet of mass $m_A$ coupling as $y_{mn} A^\mu(\bar e_m \gamma_\mu L^{\rm c}_{n})+\text{h.c.}$, we can generate ${\cal O}_{Le}$ from \Eq{eq:O2} at tree level via the $s$-channel process:
\begin{center}\vspace{-0.1cm}
\includegraphics[height=2cm]{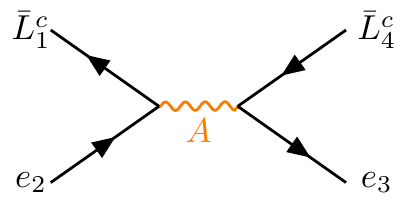}
\vspace{-0.1cm}\end{center}
We find that integrating out the massive state gives an effective operator that can be rearranged into ${\cal O}_{Le}$, with Wilson coefficients $c_{\alpha\beta}^{Le} = -|y_{mn} \alpha_n \beta_m^*|^2/m_A^2 < 0$,  in agreement with sum rule prediction in Eqs.~\eqref{eq:cbounds} and \eqref{eq:e4bound} for a vector current in the completion.
We note that, despite the presence of additional powers of momenta in the propagator numerator for the massive vector, the amplitude in this example theory is indeed perturbatively unitary, since the extra momenta are annihilated when contracted with $\gamma_\mu$ times a spinor, e.g., $[1|(p_1 + p_2)_\mu \gamma^\mu |2\rangle =  0$.
Explicitly, ${\cal A}_{\alpha\beta} = -2|y_{mn} \alpha_n \beta_m^*|^2 [13]\langle 24\rangle/(s-m_A^2 + i\epsilon)$, for which the boundary terms in the dispersion relations vanish.
Expanding in partial waves, ${\rm Im}\,a^{(j)}_{\alpha\beta} = \tfrac{1}{24} |y_{mn} \alpha_n \beta_m^*|^2 s\,\delta_{j1} \delta(s-m_A^2)$ will contribute to the right-hand side of \Eq{eq:disp}, whereas ${\rm Im}\,a_{\bar\alpha\beta}^{(j)} \propto \delta(s+m_A^2)$ will not, and we find that the sum rules are explicitly satisfied.

Having examined example $s$- and $u$-channel completions at tree level, let us finally consider a UV $t$-channel process.
For a real vector $A^\mu$ of mass $m_A$ interacting with electrons and up-type quarks via $y^e_{mn} A^\mu (\bar e_m \gamma_\mu e_n) + y^u_{mn} A^\mu (\bar u_m \gamma_\mu u_n)$, where $y^e$ and $y^u$ are both real, symmetric matrices of couplings, the amplitude for $\bar e^+ u^-$ scattering:
\begin{center}\vspace{-0.1cm}
\includegraphics[height=2cm]{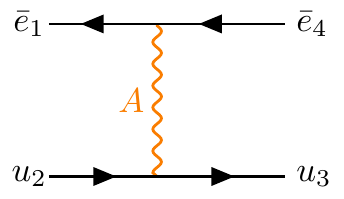}
\vspace{-0.1cm}\end{center}
is ${\cal A}_{\alpha\beta} = - 2 (y^e_{mn} \alpha_m \alpha_n^*)(y^u_{pq} \beta_p \beta_q^*)[13]\langle 24\rangle/(t{-}m_A^2 {+} i\epsilon)$,  corresponding at low energies to Wilson coefficients $c^{eu}_{\alpha\beta} = - (y^e_{mn} \alpha_m \alpha_n^*)(y^u_{pq} \beta_p \beta_q^*)/m_A^2$ of indefinite sign. 
Both ${\cal A}$ and its crossed version have imaginary parts $\propto \delta(t-m_A^2)$, which vanish for real momenta. 
While the integrals in the dispersion relations will therefore not contribute, the expressions remain valid as this is an example that contains nonzero boundary terms, $C_{\infty,\alpha\beta}^{(s)} = -C_{\infty,\alpha\beta}^{(t)} = 2c_{\alpha\beta}$, which leave the left-hand side of \Eq{eq:disp} precisely zero, matching the vanishing imaginary parts of the partial wave sum.
Thus, $t$-channel completions behave distinctly differently in the dispersion relation, in a way that shows up both in the allowed signs of Wilson coefficients and in the UV momentum scaling.

The behavior at dimension six, where the sign of the Wilson coefficient can encode information about the spin of the UV completion, stands in contrast to equivalent results at dimension eight where instead strict positivity bounds can be established.
For example, in Ref.~\cite{SMEFTfermions}, it was shown that the operator
\be
{\cal O}^{(8)}_{e} = -b^{(8),e}_{mnpq} \partial_{\mu} (\bar e_m \gamma_\nu e_n) \partial^{\mu} (\bar e_p \gamma^\nu e_q),
\ee
satisfies the positivity bound $b^{(8),e}_{mnpq} \alpha_m \beta_n \beta_p^* \alpha_q^* > 0$.
The scalar completion discussed above generates the dimension-six operator ${\cal O}_{e}$ at lowest order in $p^2/m_{\phi}^2$, and as shown, induces a positive contribution to our sum rule in \Eq{eq:e4bound}.
However, at subleading order, the theory generates ${\cal O}^{(8)}_{e}$.
Explicit computation reveals that $b^{(8),e}_{mnpq} = y_{mp} y_{nq}^*/m_{\phi}^4$, which satisfies the positivity bound.
If we consider a different UV completion, the behavior at dimension six can change, whereas that at dimension eight will not.
For instance, we could also UV-complete ${\cal O}_{e}$ using a massive vector with interaction $y_{mn} A^{\mu} (\bar{e}_m \gamma_{\mu} e_n)$, where $y_{mn} = y_{nm}^*$.
At energy scales below $m_A$, at leading order we have a contribution to the dimension-six operator from both the $s$- and $t$-channel in the UV.
By analogous arguments to those above, the $s$-channel will generate a negative contribution to our sum rule, whereas the $t$-channel will generate boundary terms and, therefore, a Wilson coefficient of indefinite sign.
Nonetheless, at next-to-leading order, we induce ${\cal O}^{(8)}_{e}$ with $b^{(8),e}_{mnpq} = y_{mn} y_{pq}/2 m_A^4$, which again satisfies the dimension-eight positivity bound.

Given that our sum rules do not specify any particular sign for individual Wilson coefficients, a single measurement is insufficient to identify whether the relations are satisfied or not.
Instead, it is from the correlations between Wilson coefficients that the generic UV properties can be inferred.
As in Ref.~\cite{SMEFTfermions}, of particular interest is the predicted connection between operators that generate experimentally distinct signatures.
In particular, the sum rule in \Eq{eq:e4bound}---and analogously for the combinations in Eqs.~\eqref{eq:abounds} and \eqref{eq:cbounds}---imply that various flavor- and CP-violating operators have coefficients with magnitudes upper-bounded by their symmetry-preserving cousins, albeit in a more complicated fashion than at dimension eight.
(As in the sum rules themselves, any such symmetry bounds will be subject to the caveats about boundary terms that we have discussed.)
Unlike at dimension eight, it is possible for a flavor-violating operator, say $c_{1123}^{eu}$, to be arbitrarily larger than $\sqrt{|c_{1122}^{eu} c_{1133}^{eu}|}$, since the presence of scalars and vectors can conspire to make one of the flavor-conserving terms vanish, due to the indefinite sign in \Eq{eq:e4bound}.
However, we can lift our definite-superposition scattering of \Eq{eq:e4bound} to a genereralized optical theorem dispersion relation (as introduced at dimension eight in Ref.~\cite{Zhang:2020jyn}).
In the case of elastic scattering, we found that when boundary terms vanish,  the partial wave expansion contains only the $j=0$ and $1$ terms for same- and opposite-helicity scattering, respectively.
It is therefore of interest to consider the scenario in which the UV completion for the general scattering process contains only scalars and vectors in these channels, in which case we find the dispersion relation:
\be 
c_{mnpq}^{eu}=8\!\int_{s_{0}}^{\infty}\frac{{\rm d}s}{s^{2}}\sum_M\!\left( M_{nq}^{(0)}M_{mp}^{(0)*} {-} 3 M_{mq}^{(1)}M_{np}^{(1)*}\right)\!,
\ee
where $M_{mn}^{(0)}$ and $M_{mn}^{(1)}$ are amplitudes for scattering $e_{m}^{+}u_{n}^{+}$ to some massive scalar and $\bar{e}_{m}^{-}u_{n}^{+}$ to a massive vector, respectively, and the sum $\sum_M$ is over all such channels in the UV.
We can treat $\int_{s_0}^\infty \frac{{\rm d} s}{s^2} \sum_M$ as an inner product, with a given scattering process viewed as a vector indexed by $s$ and the UV state $M$.
Let us write the $j=0,1$ contributions to the flavor-conserving terms as $8|\vec{M}_{12}^{(0)}|^2  = c_{1122}^{eu(0)}$, $24|\vec{M}_{12}^{(1)}|^2  = c_{1122}^{eu(1)}$, etc.
Then the flavor conserving coefficients are simply differences of their scalar and vector contributions, $c^{eu}_{1122} = c_{1122}^{(0)} - c_{1122}^{(1)}$ and $c^{eu}_{1133}= c_{1133}^{(0)} - c_{1133}^{(1)}$, while the flavor violating term is given by a difference of vector products, $c^{eu}_{1123} = 8\vec{M}_{13}^{(0)}\cdot \vec{M}_{12}^{(0)*}-24\vec{M}_{13}^{(1)}\cdot \vec{M}_{12}^{(1)*}$.
By the Cauchy-Schwarz inequality, we then have
\be
|c_{1123}^{eu}| \leq \sqrt{c_{1122}^{(0)} c_{1133}^{(0)}} + \sqrt{c_{1122}^{(1)} c_{1133}^{(1)}},\label{eq:ineq}
\ee
and similarly for all the other combinations of operators appearing in our bounds.
The individual components $c_{1122}^{(0)}$ and $c_{1122}^{(1)}$ go like $1/\Lambda^2$, where $\Lambda$ is the scale of new physics (e.g., the mass of the UV states).
If an individual Wilson coefficient were tuned to be arbitrarily small by balancing competing terms in \Eq{eq:e4bound}, naive EFT power-counting would lead to an artificially-high estimate ${\gg}\, \Lambda$ for the scale of new physics. 
However, using \Eq{eq:ineq}, if we write the flavor-violating coupling in terms of its  cutoff $\tilde \Lambda$, we have $\tilde{\Lambda} > \Lambda$ as a consequence of analyticity and unitarity when boundary terms are negligible.
Our sum rules thus offer a connection between qualitatively different types of experimental signals that would result from the leading deviations to the SM, for example, connecting collider bounds with precision decay measurements.
Violation of our sum rule would immediately reveal a property of the UV.
The fundamental field theoretic axioms used to derive our dispersion relations would imply that either boundary terms must be present (which itself tells us about the UV momentum scaling of the amplitude) or that the UV theory contains a vector and scalar tuned to cancel each other's flavor-conserving effects.

This scenario could be realized experimentally as follows.
Mu3e is targeting the range $10^{-16}$--$10^{-12}$ for ${\rm Br}(\mu\rightarrow 3e)\sim (m_W/\tilde\Lambda)^4$, which would correspond to $\tilde\Lambda$ between $80$ and $800$~TeV~\cite{Blondel:2013ia}, scales where the flavor-conserving analogues would be challenging to confirm.
Bounds on flavor-violating $\tau$ decays are much weaker, e.g., ${\rm Br}(\tau \rightarrow 3\mu) \lesssim 2\times 10^{-8}$~\cite{Zyla:2020zbs}, corresponding to $\tilde \Lambda \gtrsim 7$~TeV. 
The Belle~II experiment at SuperKEKB aims to tighten the bound on this branching ratio to $\sim 10^{-10}$, i.e., $\tilde\Lambda\sim 25$~TeV~\cite{Inami:2016aba,Kou:2018nap}.
Analogous flavor- and CP-conserving four-lepton terms have been bounded by LHC measurements~\cite{Falkowski:2015krw,Aaboud:2017yyg,Sirunyan:2018wnk,Sirunyan:2018ipj,Aad:2019fac} to $\Lambda \gtrsim 1$ to $2$~TeV.
Thus, a near-term detection of flavor violation in $\tau$ decays would imply---by virtue of our $\tilde\Lambda>\Lambda$ bound---either the near-term detection of new physics at colliders in the $10$~TeV range or else that the UV contains one of the loopholes to our arguments discussed above, in addition to being unambiguous evidence of new physics~\cite{Hernandez-Tome:2018fbq}.
Meanwhile, near-term detection of $\mu\rightarrow 3e$ could give a compelling reason to look for new lepton physics at a precision electroweak machine like the ILC and/or more broadly investigate the $100$ TeV range at a future circular collider~\footnote{We focus here on the leptonic flavor-violating implications of our bounds, since for colored fermions flavor-changing neutral currents already constrain $\tilde\Lambda \gtrsim$~PeV~\cite{Isidori:2010kg,Isidori:2013ez}, beyond the reach of planned colliders.}.

\bigskip

\begin{acknowledgments}
We thank Cliff Cheung and Nathaniel Craig for feedback and discussions, and we also thank the referees for useful comments on an earlier version of this paper.
G.N.R. is supported at the Kavli Institute for Theoretical Physics by the Simons Foundation (Grant~No.~216179) and the National Science Foundation (Grant~No.~NSF PHY-1748958) and at the University of California, Santa Barbara by the Fundamental Physics Fellowship.
N.L.R. is supported by the Miller Institute for Basic Research in Science at the University of California, Berkeley.
\end{acknowledgments}

\bibliographystyle{utphys-modified}
\bibliography{fermionsdim6}

\providecommand{\href}[2]{#2}\begingroup\raggedright\begin{thebibliography}{10}

\bibitem{Weinberg:1979sa}
S.~Weinberg, ``{Baryon and Lepton Nonconserving Processes},''
  \href{http://dx.doi.org/10.1103/PhysRevLett.43.1566}{{\em Phys. Rev. Lett.}
  {\bfseries 43} (1979) 1566}.

\bibitem{WarsawBasis}
B.~Grzadkowski, M.~Iskrzynski, M.~Misiak, and J.~Rosiek, ``{Dimension-Six Terms
  in the Standard Model Lagrangian},''
  \href{http://dx.doi.org/10.1007/JHEP10(2010)085}{{\em JHEP} {\bfseries 10}
  (2010) 085}, \href{http://arxiv.org/abs/1008.4884}{{\ttfamily arXiv:1008.4884
  [hep-ph]}}.

\bibitem{Alonso:2013hga}
R.~Alonso, E.~E. Jenkins, A.~V. Manohar, and M.~Trott, ``{Renormalization Group
  Evolution of the Standard Model Dimension Six Operators III: Gauge Coupling
  Dependence and Phenomenology},''
  \href{http://dx.doi.org/10.1007/JHEP04(2014)159}{{\em JHEP} {\bfseries 04}
  (2014) 159}, \href{http://arxiv.org/abs/1312.2014}{{\ttfamily arXiv:1312.2014
  [hep-ph]}}.

\bibitem{Lehman:2014jma}
L.~Lehman, ``{Extending the Standard Model Effective Field Theory with the
  Complete Set of Dimension-7 Operators},''
  \href{http://dx.doi.org/10.1103/PhysRevD.90.125023}{{\em Phys. Rev. D}
  {\bfseries 90} (2014) 125023},
  \href{http://arxiv.org/abs/1410.4193}{{\ttfamily arXiv:1410.4193 [hep-ph]}}.

\bibitem{Lehman:2015coa}
L.~Lehman and A.~Martin, ``{Low-derivative operators of the Standard Model
  effective field theory via Hilbert series methods},''
  \href{http://dx.doi.org/10.1007/JHEP02(2016)081}{{\em JHEP} {\bfseries 02}
  (2016) 081}, \href{http://arxiv.org/abs/1510.00372}{{\ttfamily
  arXiv:1510.00372 [hep-ph]}}.

\bibitem{Henning:2015alf}
B.~Henning, X.~Lu, T.~Melia, and H.~Murayama, ``{2, 84, 30, 993, 560, 15456,
  11962, 261485, ...: Higher dimension operators in the SM EFT},''
  \href{http://dx.doi.org/10.1007/JHEP08(2017)016}{{\em JHEP} {\bfseries 08}
  (2017) 016}, \href{http://arxiv.org/abs/1512.03433}{{\ttfamily
  arXiv:1512.03433 [hep-ph]}}.
[Erratum: \href{https://doi.org/10.1007/JHEP09(2019)019}{{\it JHEP} {\bf 09}
  (2019) 019}].

\bibitem{Murphy:2020rsh}
C.~W. Murphy, ``{Dimension-8 operators in the Standard Model Eective Field
  Theory},'' \href{http://dx.doi.org/10.1007/JHEP10(2020)174}{{\em JHEP}
  {\bfseries 10} (2020) 174}, \href{http://arxiv.org/abs/2005.00059}{{\ttfamily
  arXiv:2005.00059 [hep-ph]}}.

\bibitem{Li:2020gnx}
H.-L. Li, Z.~Ren, J.~Shu, M.-L. Xiao, J.-H. Yu, and Y.-H. Zheng, ``{Complete
  set of dimension-eight operators in the standard model effective field
  theory},'' \href{http://dx.doi.org/10.1103/PhysRevD.104.015026}{{\em Phys.
  Rev. D} {\bfseries 104} no.~1, (2021) 015026},
  \href{http://arxiv.org/abs/2005.00008}{{\ttfamily arXiv:2005.00008
  [hep-ph]}}.

\bibitem{IRUV}
A.~Adams, N.~Arkani-Hamed, S.~Dubovsky, A.~Nicolis, and R.~Rattazzi,
  ``{Causality, analyticity and an IR obstruction to UV completion},''
  \href{http://dx.doi.org/10.1088/1126-6708/2006/10/014}{{\em JHEP} {\bfseries
  10} (2006) 014}, \href{http://arxiv.org/abs/hep-th/0602178}{{\ttfamily
  arXiv:hep-th/0602178}}.

\bibitem{Pham:1985cr}
T.~N. Pham and T.~N. Truong, ``{Evaluation of the derivative quartic terms of
  the meson chiral Lagrangian from forward dispersion relations},''
\href{http://dx.doi.org/10.1103/PhysRevD.31.3027}{{\em Phys. Rev.} {\bfseries
  D31} (1985) 3027}.

\bibitem{Ananthanarayan:1994hf}
B.~Ananthanarayan, D.~Toublan, and G.~Wanders, ``{Consistency of the chiral
  pion-pion scattering amplitudes with axiomatic constraints},''
  \href{http://dx.doi.org/10.1103/PhysRevD.51.1093}{{\em Phys. Rev.} {\bfseries
  D51} (1995) 1093},
\href{http://arxiv.org/abs/hep-ph/9410302}{{\ttfamily arXiv:hep-ph/9410302
  [hep-ph]}}.

\bibitem{Pennington:1994kc}
M.~R. Pennington and J.~Portoles, ``{The chiral lagrangian parameters,
  $\ell_1$, $\ell_2$, are determined by the $\rho$-resonance},''
  \href{http://dx.doi.org/10.1016/0370-2693(94)01551-M}{{\em Phys. Lett.}
  {\bfseries B344} (1995) 399},
\href{http://arxiv.org/abs/hep-ph/9409426}{{\ttfamily arXiv:hep-ph/9409426
  [hep-ph]}}.

\bibitem{Note1}
For a broader review of the development of EFT bounds see, for example,
  Ref.~\cite {SMEFTbosons} and references therein.

\bibitem{SMEFTbosons}
G.~N. Remmen and N.~L. Rodd, ``{Consistency of the Standard Model Effective
  Field Theory},'' \href{http://dx.doi.org/10.1007/JHEP12(2019)032}{{\em JHEP}
  {\bfseries 12} (2019) 032}, \href{http://arxiv.org/abs/1908.09845}{{\ttfamily
  arXiv:1908.09845 [hep-ph]}}.

\bibitem{SMEFTfermions}
G.~N. Remmen and N.~L. Rodd, ``{Flavor Constraints from Unitarity and
  Analyticity},'' \href{http://dx.doi.org/10.1103/PhysRevLett.125.081601}{{\em
  Phys. Rev. Lett.} {\bfseries 125} (2020) 081601},
  \href{http://arxiv.org/abs/2004.02885}{{\ttfamily arXiv:2004.02885
  [hep-ph]}}.

\bibitem{Bellazzini:2018paj}
B.~Bellazzini and F.~Riva, ``{New phenomenological and theoretical perspective
  on anomalous $ZZ$ and $Z\gamma$ processes},''
  \href{http://dx.doi.org/10.1103/PhysRevD.98.095021}{{\em Phys. Rev. D}
  {\bfseries 98} (2018) 095021},
  \href{http://arxiv.org/abs/1806.09640}{{\ttfamily arXiv:1806.09640
  [hep-ph]}}.

\bibitem{Zhang:2018shp}
C.~Zhang and S.-Y. Zhou, ``{Positivity bounds on vector boson scattering at the
  LHC},'' \href{http://dx.doi.org/10.1103/PhysRevD.100.095003}{{\em Phys. Rev.
  D} {\bfseries 100} (2019) 095003},
  \href{http://arxiv.org/abs/1808.00010}{{\ttfamily arXiv:1808.00010
  [hep-ph]}}.

\bibitem{Bi:2019phv}
Q.~Bi, C.~Zhang, and S.-Y. Zhou, ``{Positivity constraints on aQGC: carving out
  the physical parameter space},''
  \href{http://dx.doi.org/10.1007/JHEP06(2019)137}{{\em JHEP} {\bfseries 06}
  (2019) 137}, \href{http://arxiv.org/abs/1902.08977}{{\ttfamily
  arXiv:1902.08977 [hep-ph]}}.

\bibitem{Wang:2020jxr}
Y.-J. Wang, F.-K. Guo, C.~Zhang, and S.-Y. Zhou, ``{Generalized positivity
  bounds on chiral perturbation theory},''
  \href{http://dx.doi.org/10.1007/JHEP07(2020)214}{{\em JHEP} {\bfseries 07}
  (2020) 214}, \href{http://arxiv.org/abs/2004.03992}{{\ttfamily
  arXiv:2004.03992 [hep-ph]}}.

\bibitem{Zhang:2020jyn}
C.~Zhang and S.-Y. Zhou, ``{Convex Geometry Perspective on the (Standard Model)
  Effective Field Theory Space},''
  \href{http://dx.doi.org/10.1103/PhysRevLett.125.201601}{{\em Phys. Rev.
  Lett.} {\bfseries 125} no.~20, (2020) 201601},
  \href{http://arxiv.org/abs/2005.03047}{{\ttfamily arXiv:2005.03047
  [hep-ph]}}.

\bibitem{Fuks:2020ujk}
B.~Fuks, Y.~Liu, C.~Zhang, and S.-Y. Zhou, ``{Positivity in electron-positron
  scattering: testing the axiomatic quantum field theory principles and probing
  the existence of UV states},''
  \href{http://dx.doi.org/10.1088/1674-1137/abcd8c}{{\em Chin. Phys. C}
  {\bfseries 45} no.~2, (2021) 023108},
  \href{http://arxiv.org/abs/2009.02212}{{\ttfamily arXiv:2009.02212
  [hep-ph]}}.

\bibitem{Yamashita:2020gtt}
K.~Yamashita, C.~Zhang, and S.-Y. Zhou, ``{Elastic positivity vs extremal
  positivity bounds in SMEFT: a case study in transversal electroweak
  gauge-boson scatterings},''
  \href{http://dx.doi.org/10.1007/JHEP01(2021)095}{{\em JHEP} {\bfseries 01}
  (2021) 095}, \href{http://arxiv.org/abs/2009.04490}{{\ttfamily
  arXiv:2009.04490 [hep-ph]}}.

\bibitem{Azatov:2021ygj}
A.~Azatov, D.~Ghosh, and A.~H. Singh, ``{Four-fermion operators at dimension 6:
  dispersion relations and UV completions},''
  \href{http://arxiv.org/abs/2112.02302}{{\ttfamily arXiv:2112.02302
  [hep-ph]}}.

\bibitem{Donal}
A.~Adams, A.~Jenkins, and D.~O'Connell, ``{Signs of analyticity in fermion
  scattering},'' \href{http://arxiv.org/abs/0802.4081}{{\ttfamily
  arXiv:0802.4081 [hep-ph]}}.

\bibitem{Davighi:2021osh}
J.~Davighi, S.~Melville, and T.~You, ``{Natural Selection Rules: New Positivity
  Bounds for Massive Spinning Particles},''
  \href{http://arxiv.org/abs/2108.06334}{{\ttfamily arXiv:2108.06334
  [hep-th]}}.

\bibitem{Gu:2020thj}
J.~Gu and L.-T. Wang, ``{Sum Rules in the Standard Model Effective Field Theory
  from Helicity Amplitudes},''
  \href{http://dx.doi.org/10.1007/JHEP03(2021)149}{{\em JHEP} {\bfseries 03}
  (2021) 149}, \href{http://arxiv.org/abs/2008.07551}{{\ttfamily
  arXiv:2008.07551 [hep-ph]}}.

\bibitem{Nicolis:2009qm}
A.~Nicolis, R.~Rattazzi, and E.~Trincherini, ``{Energy's and amplitudes'
  positivity},'' \href{http://dx.doi.org/10.1007/JHEP05(2010)095}{{\em JHEP}
  {\bfseries 05} (2010) 095}, \href{http://arxiv.org/abs/0912.4258}{{\ttfamily
  arXiv:0912.4258 [hep-th]}}. [Erratum:
  \href{https://doi.org/10.1007/JHEP11(2011)128}{{\it JHEP} {\bf 11} (2011)
  128}].

\bibitem{Froissart:1961ux}
M.~Froissart, ``{Asymptotic behavior and subtractions in the Mandelstam
  representation},''
\href{http://dx.doi.org/10.1103/PhysRev.123.1053}{{\em Phys. Rev.} {\bfseries
  123} (1961) 1053}.

\bibitem{Martin:1962rt}
A.~Martin, ``{Unitarity and high-energy behavior of scattering amplitudes},''
\href{http://dx.doi.org/10.1103/PhysRev.129.1432}{{\em Phys. Rev.} {\bfseries
  129} (1963) 1432}.

\bibitem{Bellazzini:2016xrt}
B.~Bellazzini, ``{Softness and amplitudes' positivity for spinning
  particles},'' \href{http://dx.doi.org/10.1007/JHEP02(2017)034}{{\em JHEP}
  {\bfseries 02} (2017) 034}, \href{http://arxiv.org/abs/1605.06111}{{\ttfamily
  arXiv:1605.06111 [hep-th]}}.

\bibitem{Low:2009di}
I.~Low, R.~Rattazzi, and A.~Vichi, ``{Theoretical Constraints on the Higgs
  Effective Couplings},'' \href{http://dx.doi.org/10.1007/JHEP04(2010)126}{{\em
  JHEP} {\bfseries 04} (2010) 126},
  \href{http://arxiv.org/abs/0907.5413}{{\ttfamily arXiv:0907.5413 [hep-ph]}}.

\bibitem{Falkowski:2012vh}
A.~Falkowski, S.~Rychkov, and A.~Urbano, ``{What if the Higgs couplings to W
  and Z bosons are larger than in the Standard Model?},''
  \href{http://dx.doi.org/10.1007/JHEP04(2012)073}{{\em JHEP} {\bfseries 04}
  (2012) 073}, \href{http://arxiv.org/abs/1202.1532}{{\ttfamily arXiv:1202.1532
  [hep-ph]}}.

\bibitem{Bellazzini:2014waa}
B.~Bellazzini, L.~Martucci, and R.~Torre, ``{Symmetries, Sum Rules and
  Constraints on Effective Field Theories},''
  \href{http://dx.doi.org/10.1007/JHEP09(2014)100}{{\em JHEP} {\bfseries 09}
  (2014) 100}, \href{http://arxiv.org/abs/1405.2960}{{\ttfamily arXiv:1405.2960
  [hep-th]}}.

\bibitem{Ema:2018jgc}
Y.~Ema, R.~Kitano, and T.~Terada, ``{Unitarity constraint on the K\"ahler
  curvature},'' \href{http://dx.doi.org/10.1007/JHEP09(2018)075}{{\em JHEP}
  {\bfseries 09} (2018) 075}, \href{http://arxiv.org/abs/1807.06940}{{\ttfamily
  arXiv:1807.06940 [hep-th]}}.

\bibitem{Note2}
In contrast, such a procedure would not be valid to remove the $t$-channel
  singularity in gravity, unless additional assumptions were invoked, because,
  unlike quantum electrodynamics, the EFT of general relativity is
  nonrenormalizable on its own, and thus will exhibit unitarizing UV states in
  the complex plane.

\bibitem{deRham:2017imi}
C.~de~Rham, S.~Melville, A.~J. Tolley, and S.-Y. Zhou, ``{Massive Galileon
  Positivity Bounds},'' \href{http://dx.doi.org/10.1007/JHEP09(2017)072}{{\em
  JHEP} {\bfseries 09} (2017) 072},
  \href{http://arxiv.org/abs/1702.08577}{{\ttfamily arXiv:1702.08577
  [hep-th]}}.

\bibitem{Note3}
The Weinberg operator or cubic boson operators like $W_{\mu \nu }^3$ can only
  contribute at loop level, while in our choice of basis~\cite {WarsawBasis}
  there are neither dimension-six SMEFT operators containing two bosons nor one
  boson and two fermions.

\bibitem{Note4}
We can apply this dispersion relation when $\protect \qopname \relax
  o{cos}\theta $ lies within the Martin-Lehmann ellipse; see Refs.~\cite
  {Lehmann:1958ita,Martin:1965jj,Nicolis:2009qm}. When massless loops are
  included, the ellipse is just the real interval $[-1,1]$, corresponding to
  physical $t=s(\protect \qopname \relax o{cos}\theta -1)/2$. This requirement
  poses no problem for our argument, since the contour at the origin extracting
  ${\protect \cal A}(0,t)$ can instead be taken at small, finite $s$ well below
  the UV cutoff, allowing $t$ to be finite and negative; in any case, all of
  our dispersion relations ultimately take the forward limit $t\rightarrow 0$.

\bibitem{EH}
H.~Elvang and Y.-t. Huang, {\em Scattering Amplitudes in Gauge Theory and
  Gravity}.
\newblock Cambridge University Press, 2015.
\newblock \href{http://arxiv.org/abs/1308.1697}{{\ttfamily arXiv:1308.1697
  [hep-th]}}.

\bibitem{Jacob:1959at}
M.~Jacob and G.~Wick, ``{On the General Theory of Collisions for Particles with
  Spin},'' \href{http://dx.doi.org/10.1016/0003-4916(59)90051-X}{{\em Annals
  Phys.} {\bfseries 7} (1959) 404}.

\bibitem{Horejsi:1993hz}
J.~Ho\v{r}ej\v{s}\'i, \href{http://dx.doi.org/10.1142/2445}{{\em {Introduction
  to Electroweak Unification: Standard Model from Tree Unitarity}}}.
\newblock World Scientific, 1993.

\bibitem{ItzyksonZuber}
C.~Itzykson and J.-B. Zuber, {\em Quantum Field Theory}.
\newblock McGraw-Hill, 1980.

\bibitem{Arkani-Hamed:2017jhn}
N.~Arkani-Hamed, T.-C. Huang, and Y.-t. Huang, ``{Scattering amplitudes for all
  masses and spins},'' \href{http://dx.doi.org/10.1007/JHEP11(2021)070}{{\em
  JHEP} {\bfseries 11} (2021) 070},
  \href{http://arxiv.org/abs/1709.04891}{{\ttfamily arXiv:1709.04891
  [hep-th]}}.

\bibitem{Alioli:2020kez}
S.~Alioli, R.~Boughezal, E.~Mereghetti, and F.~Petriello, ``{Novel angular
  dependence in Drell-Yan lepton production via dimension-8 operators},''
  \href{http://dx.doi.org/10.1016/j.physletb.2020.135703}{{\em Phys. Lett. B}
  {\bfseries 809} (2020) 135703},
  \href{http://arxiv.org/abs/2003.11615}{{\ttfamily arXiv:2003.11615
  [hep-ph]}}.

\bibitem{Wigner}
E.~Wigner, {\em {Gruppentheorie und ihre Anwendung auf die Quantenmechanik der
  Atomspektren}}.
\newblock Friedr. Vieweg \& Sohn Akt.-Ges., Braunschweig, 1931.
\newblock {Translated by J. J. Griffin, {\it Group Theory and its Application
  to the Quantum Mechanics of Atomic Spectra}. Academic Press, New York, 1959.}

\bibitem{Note5}
More generally, the partial wave expansion in the total angular momentum basis
  is performed with the Wigner $D$-matrices, irreducible representations of
  ${\protect \rm SU}(2)$, but these only differ from the small $d$-matrices by
  a little group phase factor, which can be set to unity by an appropriate
  choice of coordinates.

\bibitem{upcoming}
G.~N. Remmen and N.~L. Rodd, {\em {\rm to appear}}.

\bibitem{Note6}
Care must be taken not just for the relative sign between channels, but for the
  overall sign of the amplitude. The relative order of the initial and final
  two-particle states can be chosen freely, but it is convenient to define the
  final state such that in the elastic forward limit we have ${\left \langle f
  \right |}= {\left | i \right \rangle }^{\dagger }$, as this allows for a
  straightforward application of the optical theorem. To be explicit, the
  conventions we establish in the text imply that when scattering
  opposite-helicity states, we take ${\left | i \right \rangle } = {\left | 1,2
  \right \rangle }$ and ${\left \langle f \right |} = {\left \langle 3,4 \right
  |} = -{\left \langle 4,3 \right |}$. With the external states as defined,
  before performing any Wick contractions, the amplitude takes the following
  form (suppressing flavor): \begin {equation*} \begin {aligned} {\langle } 0
  {\vert } a_{p_3} b_{p_4} (\protect \bar {e} \gamma ^{\mu } e) (\protect \bar
  {e} \gamma _{\mu } e) b^{\dagger }_{p_1} a^{\dagger }_{p_2} {\vert } 0
  {\rangle }, \end {aligned} \end {equation*} where $a^{\dagger }$ and
  $b^{\dagger }$ are the particles' and antiparticles' creation operators,
  respectively. There are four possible contractions. The contraction
  $(\protect \bar {e} \gamma ^{\mu } e) (\protect \bar {e} \gamma _{\mu } e)
  \to (\protect \bar {1} 2)(\protect \bar {3} 4)$ generates no relative minus
  sign, nor does $(\protect \bar {3} 4) (\protect \bar {1} 2)$. The two
  contractions $(\protect \bar {3} 2)(\protect \bar {1} 4)$ and $(\protect \bar
  {1} 4)(\protect \bar {3} 2)$, however, will. Consequently, the full amplitude
  is given by \begin {equation*} \begin {aligned} &\protect \hspace
  {5mm}&&{\protect \cal A}(\protect \bar {e}^{-}e^{+}\to e^{+}\protect \bar
  {e}^{-}) \\&&&= -2\alpha _{m}\beta _{n}\gamma _{p}^*\delta
  _{q}^*b^e_{mnpq}({[1 \vert \gamma _{\mu } \vert 2\rangle } {[3 \vert \gamma
  ^{\mu } \vert 4\rangle } \\&&&\protect \hspace {32mm} - {[3 \vert \gamma
  _{\mu } \vert 2\rangle } {[1 \vert \gamma ^{\mu } \vert 4\rangle })
  \\&&&=-8\alpha _{m}\beta _{n}\gamma _{p}^*\delta _{q}^*b^e_{mnpq} {[13]}
  {\langle 24\rangle }, \end {aligned} \end {equation*} as given in Eq.~(\ref
  {eq:e4amp}).

\bibitem{Note7}
A similar structure appears in the computation of the nucleon-nucleon
  scattering length in the ${}^1\protect \tmspace -\thinmuskip {.1667em} S_0$
  singlet state with the operators $C_S(N^\dagger N)^2$ and $C_T (N^\dagger
  \protect \boldsymbol {\sigma } N)^2$ replacing the scalar and vector
  channels, respectively~\cite {Mehen:1999qs,Donal}.

\bibitem{Note8}
Note that the sign difference between operators of identical and mixed
  chirality does not persist to the equivalent dimension-eight terms, which
  were considered in Ref.~\cite {SMEFTfermions}. As an explicit example, in the
  notation of Ref.~\cite {SMEFTfermions}, up to a total derivative and terms
  that vanish on-shell, we have \begin {equation*} \begin {aligned} {\protect
  \cal O}_{K1}[e,L] &= - a_{mnpq}^{eL,1} (\protect \bar {e}_m \gamma _{\mu }
  \partial _{\nu } e_n) (\protect \bar {L}_p \gamma ^{\nu } \partial ^{\mu }
  L_q) \\ &= + a_{mnpq}^{eL,1} (\protect \bar {e}_m \gamma _{\mu } \partial
  _{\nu } e_n) (\partial ^{\mu } \protect \bar {L}_q^c \gamma ^{\nu } L^c_p) \\
  &= - a_{mnpq}^{eL,1} (\protect \bar {e}_m \gamma _{\mu } \partial _{\nu }
  e_n) (\protect \bar {L}_q^c \gamma ^{\nu } \partial ^{\mu } L^c_p), \end
  {aligned} \end {equation*} so that when the operator is written in terms of a
  single chirality, the same overall sign applies.

\bibitem{deBlas:2017xtg}
J.~de~Blas, J.~Criado, M.~P\'{e}rez-Victoria, and J.~Santiago, ``{Effective
  description of general extensions of the Standard Model: the complete
  tree-level dictionary},''
  \href{http://dx.doi.org/10.1007/JHEP03(2018)109}{{\em JHEP} {\bfseries 03}
  (2018) 109}, \href{http://arxiv.org/abs/1711.10391}{{\ttfamily
  arXiv:1711.10391 [hep-ph]}}.

\bibitem{Blondel:2013ia}
A.~Blondel { et~al.}, ``{Research Proposal for an Experiment to Search for the
  Decay $\mu \to eee$},''
\href{http://arxiv.org/abs/1301.6113}{{\ttfamily arXiv:1301.6113
  [physics.ins-det]}}.

\bibitem{Zyla:2020zbs}
{\bfseries Particle Data Group} {\bfseries Collaboration}, P.~Zyla { et~al.},
  ``{Review of Particle Physics},''
  \href{http://dx.doi.org/10.1093/ptep/ptaa104}{{\em PTEP} {\bfseries 2020}
  (2020) 083C01}.

\bibitem{Inami:2016aba}
{\bfseries Belle/Belle II} {\bfseries Collaboration}, K.~Inami,
  \href{http://dx.doi.org/10.22323/1.282.0574}{``{Lepton-flavor-violating
  $\tau$ decay prospects at SuperKEKB/Belle~II},''} in {\em {38th International
  Conference on High Energy Physics (ICHEP2016)}}, p.~574.
\newblock PoS, 2016.

\bibitem{Kou:2018nap}
{\bfseries Belle II} {\bfseries Collaboration}, W.~Altmannshofer { et~al.},
  ``{The Belle II Physics Book},''
  \href{http://dx.doi.org/10.1093/ptep/ptz106}{{\em PTEP} {\bfseries 2019}
  (2019) 123C01}, \href{http://arxiv.org/abs/1808.10567}{{\ttfamily
  arXiv:1808.10567 [hep-ex]}}. [Erratum:
  \href{https://doi.org/10.1093/ptep/ptz106}{{\it PTEP} {\bf 2020} (2020)
  029201}].

\bibitem{Falkowski:2015krw}
A.~Falkowski and K.~Mimouni, ``{Model independent constraints on four-lepton
  operators},'' \href{http://dx.doi.org/10.1007/JHEP02(2016)086}{{\em JHEP}
  {\bfseries 02} (2016) 086}, \href{http://arxiv.org/abs/1511.07434}{{\ttfamily
  arXiv:1511.07434 [hep-ph]}}.

\bibitem{Aaboud:2017yyg}
{\bfseries ATLAS} {\bfseries Collaboration}, M.~Aaboud { et~al.}, ``{Search for
  new phenomena in high-mass diphoton final states using 37 fb$^{-1}$ of
  proton--proton collisions collected at $\sqrt{s}=13$ TeV with the ATLAS
  detector},'' \href{http://dx.doi.org/10.1016/j.physletb.2017.10.039}{{\em
  Phys. Lett.} {\bfseries B775} (2017) 105},
\href{http://arxiv.org/abs/1707.04147}{{\ttfamily arXiv:1707.04147 [hep-ex]}}.

\bibitem{Sirunyan:2018wnk}
{\bfseries CMS} {\bfseries Collaboration}, A.~M. Sirunyan { et~al.}, ``{Search
  for physics beyond the standard model in high-mass diphoton events from
  proton-proton collisions at $\sqrt{s} = 13$ TeV},''
  \href{http://dx.doi.org/10.1103/PhysRevD.98.092001}{{\em Phys. Rev.}
  {\bfseries D98} (2018) 092001},
\href{http://arxiv.org/abs/1809.00327}{{\ttfamily arXiv:1809.00327 [hep-ex]}}.

\bibitem{Sirunyan:2018ipj}
{\bfseries CMS} {\bfseries Collaboration}, A.~M. Sirunyan { et~al.}, ``{Search
  for contact interactions and large extra dimensions in the dilepton mass
  spectra from proton-proton collisions at $\sqrt{s} = 13$ TeV},''
  \href{http://dx.doi.org/10.1007/JHEP04(2019)114}{{\em JHEP} {\bfseries 04}
  (2019) 114},
\href{http://arxiv.org/abs/1812.10443}{{\ttfamily arXiv:1812.10443 [hep-ex]}}.

\bibitem{Aad:2019fac}
{\bfseries ATLAS} {\bfseries Collaboration}, G.~Aad { et~al.}, ``{Search for
  high-mass dilepton resonances using 139 fb$^{-1}$ of $pp$ collision data
  collected at $\sqrt{s}=13$ TeV with the ATLAS detector},''
  \href{http://dx.doi.org/10.1016/j.physletb.2019.07.016}{{\em Phys. Lett.}
  {\bfseries B796} (2019) 68},
\href{http://arxiv.org/abs/1903.06248}{{\ttfamily arXiv:1903.06248 [hep-ex]}}.

\bibitem{Hernandez-Tome:2018fbq}
G.~Hern\'andez-Tom\'e, G.~L\'opez~Castro, and P.~Roig, ``{Flavor violating
  leptonic decays of $\tau$ and $\mu$ leptons in the Standard Model with
  massive neutrinos},''
  \href{http://dx.doi.org/10.1140/epjc/s10052-019-6563-4}{{\em Eur. Phys. J. C}
  {\bfseries 79} (2019) 84}, \href{http://arxiv.org/abs/1807.06050}{{\ttfamily
  arXiv:1807.06050 [hep-ph]}}. [Erratum:
  \href{https://doi.org/10.1140/epjc/s10052-020-7935-5}{{\it Eur. Phys. J. C}
  {\bf 80} (2020) 438}].

\bibitem{Note9}
We focus here on the leptonic flavor-violating implications of our bounds,
  since for colored fermions flavor-changing neutral currents already constrain
  $\protect \tilde \Lambda \gtrsim $~PeV~\cite {Isidori:2010kg,Isidori:2013ez},
  beyond the reach of planned colliders.

\bibitem{Lehmann:1958ita}
H.~Lehmann, ``{Analytic properties of scattering amplitudes as functions of
  momentum transfer},'' \href{http://dx.doi.org/10.1007/bf02859794}{{\em Nuovo
  Cim.} {\bfseries 10} (1958) 579}.

\bibitem{Martin:1965jj}
A.~Martin, ``{Extension of the axiomatic analyticity domain of scattering
  amplitudes by unitarity. 1.},''
  \href{http://dx.doi.org/10.1007/BF02720568}{{\em Nuovo Cim. A} {\bfseries 42}
  (1965) 930}.

\bibitem{Mehen:1999qs}
T.~Mehen, I.~W. Stewart, and M.~B. Wise, ``{Wigner symmetry in the limit of
  large scattering lengths},''
  \href{http://dx.doi.org/10.1103/PhysRevLett.83.931}{{\em Phys. Rev. Lett.}
  {\bfseries 83} (1999) 931},
  \href{http://arxiv.org/abs/hep-ph/9902370}{{\ttfamily arXiv:hep-ph/9902370}}.

\bibitem{Isidori:2010kg}
G.~Isidori, Y.~Nir, and G.~Perez, ``{Flavor Physics Constraints for Physics
  Beyond the Standard Model},''
  \href{http://dx.doi.org/10.1146/annurev.nucl.012809.104534}{{\em Ann. Rev.
  Nucl. Part. Sci.} {\bfseries 60} (2010) 355},
  \href{http://arxiv.org/abs/1002.0900}{{\ttfamily arXiv:1002.0900 [hep-ph]}}.

\bibitem{Isidori:2013ez}
G.~Isidori, \href{http://dx.doi.org/10.5170/CERN-2014-008.69}{``{Flavor physics
  and CP violation},''} in {\em {2012 European School of High-Energy Physics}},
  p.~69.
\newblock 2014.
\newblock \href{http://arxiv.org/abs/1302.0661}{{\ttfamily arXiv:1302.0661
  [hep-ph]}}.

\end{thebibliography}\endgroup

\end{document}